\documentclass[preprint,amsmath,amssymb,aps,prd,showpacs]{revtex4-1}

\usepackage{float}

\usepackage{epsfig}
\usepackage{dcolumn}
\usepackage{bm}
\usepackage{tikz}
\usepackage{graphicx, graphics, color}
\usepackage{dcolumn}
\usepackage{bm}
\usepackage{hyperref}
\usepackage[mathlines]{lineno}

\newcommand{\be}{\begin{equation}}
\newcommand{\ee}{\end{equation}}
\newcommand{\ben}{\begin{eqnarray}}
\newcommand{\een}{\end{eqnarray}}

\newcommand{\cK}{{\cal K}}

\newcommand{\cO}{{\cal O}}

\newcommand{\cL}{{\cal L}}

\newcommand{\cM}{{\cal M}}

\newcommand{\cB}{{\cal B}}
\newcommand{\cU}{{\cal U}}

\newcommand{\p}{\partial}
\newcommand{\na}{\nabla}

\newcommand{\hSi}{{\hat \Sigma}}

\newcommand{\hg}{\hat g}

\newcommand{\tg}{\tilde g}

\newcommand{\ep}{\epsilon}

\newcommand{\ga}{\gamma}

\newcommand{\tR}{{\tilde R}}

\newcommand{\tna}{\tilde \na}

\begin{document}

\title{Uniqueness of electrically and magnetically charged phantom wormholes}
\author{Marek Rogatko} 
\email{rogat@kft.umcs.lublin.pl}
\affiliation{Institute of Physics, 
Maria Curie-Sklodowska University, 
20-031 Lublin, pl.~Marii Curie-Sklodowskiej 1, Poland}

\date{\today}

\begin{abstract}
We prove the uniqueness theorem  for static spherically symmetric traversable wormholes with two asymptotically flat ends, constituting the 
solutions of Einstein-phantom-electric-magnetic equations of motion. For the completeness of the results, we also consider the case of
phantom $U(1)$-gauge field. The key role in the proof is bounded with conformal positive energy theorem.
\end{abstract}

\maketitle

\section{Introduction}
The history of wormholes being the solutions of general relativity or alternative gravity theories and constituting form of bridges among different regions of spacetime
or distinct regions of Universes, goes back to \cite{ein35}, where the so-called Einstein-Rosen bridges have been proposed. They account for the bridges connecting
two exterior regions of Schwarzschild black hole and at first they comprise
geometric models of elementary particles \cite{mis57} or describe the initial data for Einstein equations \cite{mis60}. Next the model having no 
event horizon and physical singularities, was conceived in Refs. \cite{ell73}-\cite{ell79}. 

A special class of wormholes which is characterized as a region of nontrivial topology located inside a compact spatial domain, that particles and light are able 
to penetrate through a topological handle and return to the exterior region without encountering singularities, has been proposed \cite{mor88}-\cite{hoc97}.
It was also shown \cite{mor88,fri90}
that a relative motion of traversable wormholes' mouths may generate closed timelike curves. In the case when the traversable wormhole mouth is not moving, 
but one of them is surrounded by mass distribution \cite{fro90}, it leads to alike effect.

The simplest model of a traversable wormhole connecting two asymptotically flat regions has a spherically symmetric geometry, which means that there exists a two-sphere
of a minimal area in the throat and the null-energy
conditions should be violated in its vicinity. The requirement for a throat held open is to invoke phantom field (exotic matter) with reversed sign kinetic energy term,
whose energy momentum tensor violates the null energy condition.

It should be mentioned that in order to avoid the implementation of exotic matter, the modified gravity traversable wormholes have been discussed (see, e.g,  \cite{kan11}-\cite{can23}).
The necessity of the phantom filed can also be evade by the implementation of higher curvature terms in gravity actions.

The other way out is to use quantum fields in the semi-classical formulation of gravity. It happens that quantum fields may violate null energy condition \cite{eps65} and provide required amount 
of negative energy density to support the existence of traversable wormhole \cite{qua}.

On the other hand, in Refs \cite{gib16,gib17} the model with asymptotically flat regions connected through a disk, which boundary accounts for a circular ring with  a conical curvature singularity,
was proposed. The generalization of Bronikov-Ellis and vacuum ring wormholes to globally regular stationary solutions were analyzed in \cite{vol21}.
In addition much attention was paid to the stationary wormhole solutions \cite{teo98}-\cite{kle14} and it was claimed that rotating solutions would be probably stable  \cite{mat06} and therefore traversable.

Moreover, an electrically charged traversable wormhole solution in Einstein-Maxwell -phantom dilaton gravity, by the assumption that the dilaton is a phantom field, was 
obtained in \cite{gou18}, while properties of electrically charged wormholes constructed from a complex self-interacting charged field were elaborated in \cite{jar24}.
In \cite{can24} the construction of electric/magnetic generalization of Ellis-Bronikov wormhole in the framework of Einstein gravity with linear/non-linear electrodynamics and exotic dust matter
being pressure-less perfect fluid with negative energy density, was considered. 
  
It was also revealed that wormhole solutions emerge naturally from the effective action  stemming from heterotic string theory \cite{das90}.
 On the other hand, important from the point of view of the unification scheme, higher-dimensional wormhole solutions were elaborated in Refs. \cite{bro97}-\cite{tor13}.

Large amount of works was devoted to the possible astrophysical implications of the wormhole existence, e.g., the system like neutron star wormhole \cite{dzh11,dzh14}, 
wormholes inside stars and black holes \cite{noj24}, gravitational 
lensing effects caused by these objects \cite{cra95}-\cite{jus17}, behavior of the dark matter clouds \cite{kic22}.

From the above it can be seen that wormhole solutions appear in many cases (see \cite{book} and references therein), from general relativity, alternative gravity 
theories to higher generalization of Einstein theory.
Thus having in mind classification of black holes (uniqueness theorem for them) one want to conceive the systematic classification of the wormhole solutions.
The first work in this direction was provided by Ruback in \cite{rub89},
where the uniqueness theorem for wormhole spaces with vanishing Ricci scalar was studied. 
In \cite{yaz17} the uniqueness
of Ellis-Bronikov wormhole with phantom field was proposed. The higher-dimensional case of the static spherically symmetric phantom wormholes was treated in \cite{rog18a}.
In \cite{laz17} the uniqueness for four-dimensional case of the Einstein-Maxwell-dilaton wormholes with the dilaton coupling constant equal to one, was elaborated.
The uniqueness of static spherically symmetric traversable wormholes with two asymptotically flat ends, subject to the higher-dimensional solutions of Einstein-Maxwell-phantom dilaton field equations was proved in \cite{rog18b}, where the case of an arbitrary dilaton coupling constant was elaborated. 

In our paper we pay attention to the problem of the classification (uniqueness theorem) of wormholes with phantom and both electric and magnetic Maxwell fields.

Global, regular spherically symmetric solutions with scalar and electromagnetic fields which constituted the traversable wormholes were given for instance in Refs. \cite{bol12}-\cite{kar06}.

The other aspect of our motivation was connected with works considering electromagnetic wormholes in metamaterials, i.e., devices working like wormholes with respect to Maxwell equations
and effectively changing the topology of space near electromagnetic wave propagation \cite{gre07}, or the experimental realization of the aforementioned idea
(magnetostatic wormhole in metamaterials).
It was shown that such a device transferred the magnetic field from one point in space to another through a path which was magnetically undetectable.
Magnetic field from a source at one end of wormhole appeared at the other end, creating the illusion of magnetic field propagating through a tunnel outside
three-dimensional space \cite{pra15}.

On the other hand,
the influence of magnetic field on compact objects, like black holes and wormholes, seems to be important in the view of the recent measurements of
polarization conducted by the Event Horizon Telescope  (EHT) Collaboration, which reveals the signature of magnetic field close to the edge of supermassive black holes
in M87 and NGC 1052 galaxies \cite{mag1}-\cite{mag4}. On the other hand, these observations are important from the point of view of rising hope for the future verifications
of compact object characteristics and the possible verification of the wormhole existence.

After deriving equation of motion for the strictly static, asymptotically flat and simply connected spacetime, in Sec. III, we apply the conformal positive theorem \cite{sim99}. 
Examination of the boundary conditions and next showing the conformal flatness of the wormhole in question spacetime, will be crucial in the proof of the uniqueness.
For the generality of investigations one considers the case of ordinary and phantom $U(1)$-gauge fields.
In the last section we conclude our investigations.

\section{Uniqueness theorem for electrically and magnetically charged wormholes}
Our main aim will be to study traversable wormhole with both magnetic and electric charges.
The action of the studied problem will be provided by
\be
S = \int d^4 x \sqrt{-g}~ \Big( R + 2 ~\na_\mu \phi \na^\mu \phi  - \ep~ F_{\mu \nu} F^{\mu \nu} \Big),
\ee
where $R$ is the Ricci scalar in the spacetime under inspection, $\na_\alpha$ denotes Levi-Civita connection in the manifold in question.
The constant $\ep = \pm 1$, will be kept for the generality of the considerations, i.e., it allows one to examine the case of the phantom $U(1)$-gauge field.
$ F_{\mu \nu} = 2 \na_{[\mu} A_{\nu ]}$ stands for the  $U(1)$-gauge field strength.

\subsection{Equations of motion in the presence of static Killing vector field}

We shall consider the static spacetime in the strict sense, with a timelike Killing vector field $k_\alpha = (\p/\p t)_\alpha$ which will be given in each point of the manifold.
The line element of the spacetime in question is provided by
\be
ds^2 = - N^2 dt^2 + g_{ij} dx^i dx^j,
\ee
where $g_{ij}$ is the metric tensor of three-dimensional Riemannian manifold, while $N$ is a smooth lapse function. They are time independent because of the fact that
$g_{ij}$ and $N$ are defined on constant time hypersurface.

Moreover, Maxwell and phantom scalar satisfy the staticity condition of the form as follows:
\be
\cL_{k} F_{\mu \nu} = 0, \qquad \cL_k \phi = 0,
\ee
where $\cL_k $ stands for the Lie derivative with respect to the Killing vector field $k_\alpha$.

The existence of stationary Killing vector field $k_\alpha$ in the considered manifold allow us to introduce
the twist vector $\omega_\alpha$, provided by the following relation:
\be
\omega_\alpha = \frac{1}{2}~\ep_{\alpha \beta \ga \delta}~k^\beta~\na^\ga~k^\delta.
\ee
On the other hand, it can be shown that for any Killing vector field we get
$\na_\alpha~\na_\beta \chi_\ga = - R_{\beta \ga \alpha}{}{}^{\delta}~\chi_\delta$, which yields the relation
 \be
\na_\beta ~\omega_\alpha = \frac{1}{2}~\ep_{\alpha \beta \ga \delta}~k^{\ga}~R^{\delta \chi}~k_{\chi}.
\label{rrq}
\ee

Additionally, one can also check that
$
\na_\alpha~\Big( \frac{\omega^\alpha}{N^4} \Big) = 0,$ 
 where we set $N^2 = - k_\ga~k^\ga$.

Furthermore the introduction of static Killing vector field $k_\alpha$,  leads to the definitions of
electric and magnetic components for gauge field strengths $F_{\alpha \beta}$, given by
\be
E_{\alpha} = - F_{\alpha \beta}~k^\beta, \qquad
B_{\alpha} = \frac{1}{2}~\ep_{\alpha \beta \ga \delta}~k^\beta~F^{\ga \delta}.
\label{eb}
\ee
Having in mind the above equation (\ref{eb}), the field strength $F_{\alpha \delta}$ may be rewritten in terms of $E_{\alpha}$ and $B_{\alpha}$, in the form as
$N^2~F_{\alpha \beta}= - 2~k_{[\alpha} E_{\beta]}+ 
\ep_{\alpha \beta \ga \delta}~k^\ga~B^{\delta}.$

Consequently, the equations of motion
for magnetic and electric parts gauge field strength imply the following:
\ben \label{prop1}
\na_{\alpha}\bigg( {E^{\alpha} \over N^2} \bigg) &=& 2~{B^{\ga} \over N^4}~\omega_\ga,\\ \label{prop2}
\na_{\alpha}\bigg( {B^{\alpha} \over N^2} \bigg) &=& - 2~{E^{\ga} \over N^4}~\omega_\ga.
\een
On the other hand, using the conditions
$\cL_k F_{\alpha \beta}= 0$ and the
relations $\na_{[ \ga} F_{\alpha \beta ]} = 0$, implies that the generalized Maxwell source-free equations can be rewritten as
\be
\na_{[\alpha}E_{\beta ]} = 0,\qquad
\na_{[\alpha}B_{\beta ]} = 0.
\ee
Moreover if one considers the simply connected spacetime, the electric and magnetic potentials yield
\be
E_{\alpha} = \na_{\alpha} \psi_{F}, \qquad
B_{\alpha} = \na_{\alpha} \psi_B.
\label{FB}
\ee
The relations (\ref{rrq}) and the explicit form
of the Ricci tensor, enable to find the {Poynting flux} in Einstein-Maxwell gravity with electric and magnetic charges, being subject to the relation
\be
\na_{[\alpha} \omega_{\beta ]} = 4~ E_{[ \alpha} B_{\beta ]}.
\label{oom}
\ee 
Consequently, the dimensionally reduced equations of motion for Einstein-Maxwell-phantom fields are given by
\ben \label{rr}
{}^{(g)}R_{ij} &-& \frac{1}{N} {}^{(g)} \na_i {}^{(g)} \na_j N = - 2 {}^{(g)} \na_i  \phi {}^{(g)} \na_j \phi \\ \nonumber \label{np}
&-&2 \ep~\Big[ \frac{{}^{(g)} \na_i \psi_F {}^{(g)} \na_j \psi_F}{N^2} - g_{ij} \frac{ {}^{(g)} \na_k \psi_F {}^{(g)} \na^k \psi_F}{2 N^2} \Big] \\ \nonumber
&-&2 \ep~\Big[ \frac{{}^{(g)} \na_i \psi_B {}^{(g)} \na_j \psi_B}{N^2} - g_{ij} \frac{ {}^{(g)} \na_k \psi_B {}^{(g)} \na^k \psi_B}{2 N^2} \Big], \\
{}^{(g)} \na_i {}^{(g)} \na^i \phi &+& \frac{{}^{(g)} \na_i N {}^{(g)} \na^i \phi}{N} = 0, \\ 
{}^{(g)} \na_i {}^{(g)} \na^i \psi_F &-& \frac{{}^{(g)} \na_i N {}^{(g)} \na^i \psi_F}{N} =0,\\
{}^{(g)} \na_i {}^{(g)} \na^i \psi_B &-& \frac{{}^{(g)} \na_i N {}^{(g)} \na^i \psi_B}{N} =0,\\
{}^{(g)} \na_i {}^{(g)} \na^i N &-& \frac{\ep}{N} \Big( {}^{(g)} \na_k \psi_F {}^{(g)} \na^k \psi_F + {}^{(g)} \na_k \psi_B {}^{(g)} \na^k \psi_B \Big)  = 0, 
\een 
where ${}^{(g)}R_{ij}$ and ${}^{(g)} \na_i $ are Ricci scalar curvature and connection existing in three-dimensional spacetime, while $\psi_F$ and
$\psi_B$ are electrostatic and magnetic potentials, as defined by the relation (\ref{FB}).

\subsection{Properties of wormhole geometry}
In order to achieve the characteristics of the considered wormhole manifold, we suppose that the spacetime in question
will be strictly static one, which ensures that we have no event horizons. 
Moreover one assumes that
the three-dimensional submanifold with metric tensor $g_{ij}$, will be
complete, i.e., we have three-dimensional hypersurfaces $\Sigma$
of constant time which are singularity free, and
for a compact subset $\cK \subset \Sigma$, consisting of two ends $\Sigma_{\pm}$ diffeomorphic to $R^{(3)}/B^{(3)}$, where $B^{(3)}$
is closed unit ball situated at the origin of $R^{(3)}$, 
we have a standard coordinate system in which the expansion of the following forms is given:
\ben \label{a}
g_{ij} &=& \Big( 1 +  \frac{2 M_\pm}{r} \Big) \delta_{ij} + \cO \Big(\frac{1}{r^{2}}\Big),\\
N &=& N_{\pm} \bigg(1 - \frac{M_\pm}{r} \bigg) +  \cO \Big(\frac{1}{r^{2}} \Big),\\
\psi_F &=& \frac{Q_{F \pm}}{r} + \cO \Big(\frac{1}{r^{2}} \Big),\\ \label{a3}
\psi_B &=& \frac{Q_{B \pm}}{r} + \cO \Big(\frac{1}{r^{2}} \Big),\\ \label{a3b}
\phi &=& \phi_{\pm} - \frac{q_\pm}{ r} + \cO \Big(\frac{1}{r^{2}} \Big),
\een
where $N_\pm>0,~\phi_\pm,~\mu_\pm,~Q_{F,B \pm},~q_\pm$ are constant.
$M_\pm$ and $q_\pm$ represent the ADM masses and scalar charges, while the electric and magnetic charges
of the two ends $\Sigma_{\pm}$, are denoted respectively
by $Q_{F,B \pm}$. 
The above relations (\ref{a})-(\ref{a3b}) are connected with the standard notion of asymptotically flat regions.

\subsection{Uniqueness theorem}
Before we proceed to the main subject of this section, we pay attention to the relation between electric and magnetic potentials, $\psi_{F},~\psi_B$, in the case 
when simply connected, asymptotically flat, strictly static spacetime is elaborated. Namely, when we have to do with static spacetime with Killing vector field $k_\alpha$, the right-hand side of the relation
(\ref{oom}) is equal to zero, which implies that we have proportionality between electric and magnetic components of Maxwell fields. The electric one-form 
is spacelike, because of the fact that $k_\alpha$ is timelike, which caused that every one-form parallel 
and orthogonal to it, disappears (\ref{prop1})-(\ref{prop2}).

Consequently
the asymptotic conditions for the potentials
$\psi_F \rightarrow 0$
and $\psi_B \rightarrow 0$, when $r \rightarrow \infty$, lead to the conclusion that
\be
\psi_B = \mu~\psi_F,
\label{elmag}
\ee
where $\mu$ is a constant value.

In our considerations the most important will be the conformal positive energy theorem \cite{sim99}.
For the readers convenience we pay some attention to its formulation. 

First of all one ought to have two asymptotically flat 
Riemannian $(n-1)$-dimensional manifolds. say, 
$(\Sigma^{\Phi},~ {}^{(\Phi)}g_{ij})$ and $(\Sigma^{\Psi},~ {}^{(\Psi)}g_{ij})$, for which
metric tensors are connected by the conformal transformation
given by ${}^{(\Psi)}g_{ij} = \Omega^2~{}^{(\Phi)}g_{ij}$, wiith $\Omega$ conformal factor.
Further it implies that the masses of the considered
manifolds  fulfils the relation
${}^{\Phi}m + \beta~{}^{\Psi}m \geq 0$ if ${}^{(\Phi)} R + \beta~\Omega^2~{}^{(\Psi)} R \geq 0$, 
where ${}^{(\Phi)} R $ and ${}^{(\Psi)} R$ are the Ricci scalars bounded with the adequate metric tensors, while
$\beta$ is a positive constant. 
The equality is fulfilled if and only if the
$(n-1)$-dimensional manifolds are flat \cite{sim99, gib02}.


Having in mind the above assumptions to the conformal positive energy theorem \cite{sim99}, let us suppose
that three-dimensional metric tensor will be defined by the following standard conformal transformation
\be
\tg_{ij} = N^{2} ~g_{ij},
\ee
while the Ricci curvature tensor in the conformally rescaled metric will be provided by
\be  \label{rij} 
 \tR(\tg)_{ij} = \frac{2}{N^2} \tna_i N \tna_j N 
- 2 \tna_i \phi \tna_j \phi - 2 \ep \frac{(1+\mu)^2 \tna_i \psi_F \tna_j \psi_F }{N^2},
\ee
where $\tna_k$ are connected with $\tg_{ij}$.

In the next step of the proof, let us define quantities of the forms as
\ben
\Phi_{1}&=& \frac{1}{2} \Big(N + \frac{1}{2N} \Big),\\
\Phi_{-1} &=&\frac{1}{2} \Big(N - \frac{1}{2N} \Big),,\\
\Phi_{0} &=& i ~\phi,
\een
and
\ben
\Psi_{1} &=& \frac{1}{2} \Big( N + \frac{1}{2N} -2 \frac{\ep (1 +\mu^2) \psi_F^2}{N^2} \Big),\\
\Psi_{-1} &=& \frac{1}{2} \Big( N - \frac{1}{2N} -2 \frac{\ep (1 +\mu^2) \psi_F^2}{N^2} \Big),\\
\Psi_0 &=& \frac{\psi_F}{N}.
\een

On the studied spacetime the following symmetric tensor can be determined
\ben \label{ricph}
{}^{(\Phi)} \tR_{ij} &=& \tna_i \Phi_{-1} \tna_j \Phi_{-1} - \tna_i \Phi_0 \tna_j \Phi_0 
- \tna_i \Phi_1 \tna_j \Phi_1, \\ \label{ricpsi}
{}^{(\Psi)} \tR_{ij} &=& \tna_i \Psi_{-1} \tna_j \Psi_{-1} - \tna_i \Psi_0 \tna_j \Psi_0 -  \tna_i \Psi_{1} \tna_j \Psi_{1},
\een
where we have defined the metric of the form $\eta_{AB} = diag(1,-1,-1)$. Moreover one has two constraints, which are given by 
$\Psi_{B} \Psi^{B} = \Phi_{B} \Phi^B = -1$, where $B= (0,~1,-1)$. 

Having in mind the relations (\ref{ricph}) and (\ref{ricpsi}), as well as constraint relations, one arrives at the following:
\be
\tna_k \tna^k \Psi_B  = {}^{(\Psi)} \tR_{i}{}{}^{i} ~\Psi_B, \qquad
\tna_k \tna^k \Phi_B  = {}^{(\Phi)} \tR_{i}{}{}^{i} ~\Phi_B.
\label{nab}
\ee
Consequently, it can be checked by the direct calculations that the Ricci curvature tensor of the conformally rescaled metric $\tg_{ij} $ may be written as follows:
\be
\tR_{ij} = 2 \Big( {}^{(\Phi)} \tR_{ij} + {}^{(\Psi)} \tR_{ij} \Big).
\label{confr}
\ee
One points out that the relations for $\Phi_{A}$ and $\Psi_A$, as was shown for the first time in \cite{mar02, hoe76, sim92},
can be achieved from the following Lagrangian density:
\be
\cL = \sqrt{-\tg} \Big( {}^{(\Phi)} \tR_{m}{}{}^{m} + {}^{(\Psi)} \tR_{m}{}{}^{m}
+ \frac{ \tna^i  \Phi_{A} \tna_i  \Phi^{A}}{ \Phi_{A} \Phi^{A}} + \frac{ \tna^i  \Psi_{A} \tna_i  \Psi^{A}}{ \Psi_{A} \Psi^{A}} \Big),
\ee
by the independent variation with respect to $\tg_{ij},~ \Phi_{A}, ~ \Psi_{A}$, and the application of the constraint relations for $\Phi_A$ and $\Psi_A$.

\subsubsection{Conformal transformations and conformal positive energy theorem}
In this subsection we shall apply the conformal positive energy theorem \cite{sim99} to conduct the uniqueness proof  of the 
spherically symmetric asymptotically flat wormholes with phantom and both electric and magnetic Maxwell fields.
Due to the assumptions of the theory in question one has to define two Riemannian manifolds with given conformal transformations which bind their
metric tensors. On the account of these requirements, it is customary to define
the conformal transformations given by
\be
{}^{(\Phi)}g_{ij}^{\pm} = {}^{(\Phi)}\omega_{\pm}^{2}~ \tg_{ij},
\qquad
{}^{(\Psi)}g_{ij}^{\pm} = {}^{(\Psi)}\omega_{\pm}^{2}~ \tg_{ij}.
\label{pff}
\ee
Their conformal factors imply the following:
\be
{}^{(\Phi)}\omega_{\pm} = \frac{\Phi_{1} \pm 1 }{ 2}, \qquad
{}^{(\Psi)}\omega_{\pm} = \frac{\Psi_{1} \pm 1}{  2}.
\label{pf}
\ee
Further, let us implement the procedure of pasting three-dimensional manifolds \cite{gib02,gib02a}
 $(\Sigma_{\pm}^{\Phi},~ {}^{(\Phi)}g_{ij}^{\pm})$ 
and 
$(\Sigma_{\pm}^{\Psi},~ {}^{(\Psi)}g_{ij}^{\pm})$ across their shared minimal boundary.
The procedure in question enables us to get four manifolds
four manifolds $(\Sigma_{+}^{\Phi},~ {}^{(\Phi)}g_{ij}^{+})$,
$(\Sigma_{-}^{\Phi},~ {}^{(\Phi)}g_{ij}^{-})$,~ $(\Sigma_{+}^{\Psi},~ {}^{(\Psi)}g_{ij}^{+})$,~\\$(\Sigma_{-}^{\Psi},~ {}^{(\Psi)}g_{ij}^{+})$.
Those manifolds will be pasted
across shared minimal boundaries, which we denote by  ${\cB}^\Psi$ and ${\cB}^\Phi$.

This construction leads to obtaining complete regular hypersurfaces 
$\Sigma^{\Phi} = \Sigma_{+}^{\Phi} \cup \Sigma_{-}^{\Phi}$ and $\Sigma^{\Psi} = \Sigma_{+}^{\Psi} \cup \Sigma_{-}^{\Psi} $.
Consequently, completeness of 
${}^{(\Phi)}g_{ij}^{\pm}$ and ${}^{(\Phi)}g_{ij}^{\pm}$ metrics, follows from their definitions, completeness of $g_{ij}$ and
the inequalities $N_{-}  \le N \le N_+$. The last inequalities can be found using the maximum principle for elliptic partial differential equations, as well as,
the asymptotic behavior of $N$. The equality takes place when $Q_{F \pm}=Q_{B \pm} = M_{\pm} =0$.


As has been revealed in Refs. \cite{gib02,gib02a}, it can be checked that the total gravitational mass ${}^{\Phi} m$ on hypersurface $\Sigma^{\Phi}$ and ${}^{\Psi}m$
 on $\Sigma^{\Psi}$ vanish, i.e., it can shown that the metric tensors connected with the adequate hypersurfaces are proportional to Kronecker delta.

Thus, summing it all up, we conclude that 
the resulting manifolds $\Sigma^{\Phi} $ and $\Sigma^{\Psi} $ are geodesically complete, with one asymptotically flat end of vanishing total gravitational mass,
from now on we shall denote them respectively by $\Sigma^{\Phi}_+ $ and $\Sigma^{\Psi}_+$. 

In order to show the static slice is conformally flat, it is customary to define 
another conformal transformation given by
\be
{\hat g}^{\pm}_{ij} = \bigg[ \bigg( {}^{(\Phi)}\omega_{\pm} \bigg)^2
 \bigg( {}^{(\Psi)}\omega_{\pm} \bigg)^{2} \bigg]^{1 \over 2}\tg_{ij}.
\ee
Consequently it guides to form of
the Ricci curvature tensor on the considered manifold
\ben \label{ricth}
\hat R_\pm &=& \bigg[ {}^{(\Phi)}\omega_{\pm}^2~ {}^{(\Psi)}\omega_{\pm}^{2 } \bigg]
^{-{1 \over 2}}
\bigg( {}^{(\Phi)}\omega_{\pm}^{2} {}^{(\Phi)}R_\pm +
{}^{(\Psi)}\omega_{\pm}^{2} {}^{(\Psi)}R_\pm \bigg) \\ \nonumber
&+& 
\bigg( \hat \na _{i} \ln {}^{(\Phi)}\omega_{\pm} - {\hat \na} _{i} \ln {}^{(\Psi)}\omega_{\pm} \bigg)  
\bigg( \hat \na ^{i} \ln {}^{(\Phi)}\omega_{\pm} - {\hat \na}^{i} \ln {}^{(\Psi)}\omega_{\pm} \bigg).  
\een
By the direct calculations, it can be revealed that 
the first term on the right-hand side of the equation (\ref{ricth}) implies
\ben \label{pos1}
{}^{(\Phi)}\omega_{\pm}^{2}~ {}^{(\Phi)}R_\pm + {}^{(\Psi)}\omega_{\pm}^{2}~ {}^{(\Psi)}R_\pm &=& 
2~\mid {\Phi_{0} \tna_{i} \Phi_{-1}
- \Phi_{-1} \tna_{i} \Phi_{0} \over
\Phi_{1} \pm 1 } \mid^2 \\ \nonumber
&+& 2~\mid { \Psi_{0} \tna_{i} \Psi_{-1}
- \Psi_{-1} \tna_{i} \Psi_{0} \over
\Psi_{1} \pm 1} \mid^2.
\een
Inspection of the relations (\ref{ricth}) and (\ref{pos1}) reveals that the Ricci scalar $\hat R_\pm $ is greater or equal to zero.

On the other hand, from the conformal positive energy theorem \cite{sim99} one concludes that
$(\Sigma^{\Phi}_+,~ {}^{(\Phi)}g_{ij})$, $(\Sigma^{\Psi}_+,~ {}^{(\Psi)}g_{ij})$ and
$(\hSi_+,~ \hg_{ij})$ are flat. Moreover these facts yield that
the conformal factors
${}^{(\Phi)}\omega = const  {}^{(\Psi)}\omega$ and $\Phi_{1} = \Psi_{1}$, as well as,
$\Phi_{0} = const~ \Phi_{-1}$ and $\Psi_{0} = const~ \Psi_{-1}$. 

Hence, all the above provide the conclusion that the manifold $(\Sigma_+,~ g_{ij})$ is conformally flat.

This fact enables to rewrite the metric tensor $\hg_{ij}$ in a 
conformally flat form
\be
\hg_{ij} = {\cal U}^{4}~ {}^{(\Phi)}g_{ij},
\label{gg}
\ee
with the conformal factor given by ${\cal U} = ({}^{(\Phi)}\omega_{\pm} N)^{-1/2}$.

Calculating Ricci scalar $\hat R$, we get ${}^{(\Phi)}R$ plus term proportional to $\nabla^2 {\cal U}$ \cite{gib02a}. 
Because of the fact that $\hat R = {}^{(\Phi)}R = 0$, one has that
$\cal U$ is harmonic function  on the three-dimensional Euclidean manifold
$
\na_{i}\na^{i}{\cal U} = 0,
$
where by $\na$ one denotes the covariant derivative on a flat manifold. 

As in \cite{rog18a,rog18b}, one defines a local coordinate in the neighborhood $\cM \in \hSi_+$
\be
{}^{(\Phi)} g_{ij} dx^i dx^j = \delta_{ij}dx^{i}dx^{j} = {\tilde \rho}^{2} d{\cU}^2 + {\tilde h}_{AB}dx^{A}dx^{B},
\ee
where we have defined ${\tilde \rho}^2 = \na_b {\cU} \na^b {\cU}.$

The manifold in question is totally geodesic, which means that any of its sub-manifold geodesic is a geodesic in the considered manifold.
The other important fact is that the embedding of $\hSi_+$ into Euclidean three-dimensional manifold is totally umbilical \cite{kob69},
which ensues that each connected 
component of $\hSi_+$ is a geometric sphere 
of a certain radius.  Without loss of the generality, the considered embedding has the feature of being rigid \cite{kob69}. It results in the fact that
one is always able to locate one connected wormhole of a certain radius $\rho$, say at $r=r_0$ surface on $\hSi_+$,
which in turn enables us to introduce the metric tensor of the following form:
\be
\tg_{ij}dx^i dx^j = \rho^2 dN^2 + h_{AB}dx^A dx^B.
\ee

The above mathematical construction leads us to a boundary value problem for the Laplace equation on the base space $\Theta = E^{3}/B^{3}$, with a Dirichlet boundary conditions.
Our system 
is characterized by a parameter which fixes the radius of the inner boundary  and authorizes wormhole of a specific radius $\rho$. 

To proceed further, let us assume that one has two solutions of the considered system equations of motion, being subject to the same boundary value problem. Using the Green
identity and integrating over the volume $\Theta$, we arrive at the following:
\ben \label{green} \nonumber
\Big( \int_{r \rightarrow \infty} - \int_{\Sigma_{wh}} \Big) (\cU_1- \cU_2)~\frac{\p}{\p r} (\cU_1-\cU_2) dS \\
= \int_\Theta \mid \na (\cU_1 - \cU_2) \mid^2 d \Theta.
\een
We may conclude that the surface integrals on the left-hand side of the equation (\ref{green})
vanish because of the imposed boundary conditions and it provides that the volume  integral have to be identically equal to zero.
It all leads to the conclusion that the considered two solutions of the Laplace equation with the Dirichlet boundary conditions are identical. \\

\noindent
{\bf Theorem:}\\
Let us suppose that $\cU_1$ and $\cU_2$ are two solutions of the Laplace equation on the base space defined above, 
being the solutions of Einstein-phantom-electric-magnetic equations of motion of static spherically symmetric traversable wormholes  with two asymptotically flat ends.
The aforementioned solutions fulfil the same boundary and regularity conditions. Thus, one has that $\cU_1 = \cU_2$ in all of the region of the base space $\Theta$, yielded that
$\cU_1(p) = \cU_2(p)$ for at least one point in the described region.\\

\section{Conclusions}
In our paper we have considered the uniqueness theorem
(classification scheme) for the static, spherically symmetric, traversable wormhole with two asymptotically flat ends, being the solution of 
Einstein-phantom-electric-magnetic theory. For the completeness of the results one elaborates the case of ordinary and phantom $U(1)$-gauge fields.
The main tool in our considerations was the conformal positive energy theorem \cite{sim99}, and conformal transformations for the system in question, enabling us to apply the
aforementioned theorem.
The equations of motion simplify if we take into account the static spacetime, which enables us to find that on simply connected manifold, the relation binding 
electric and magnetic potentials. 

In the end we reveal that two solutions of Einstein-phantom electric-magnetic gravity, subject to the same boundary and regularity conditions, are the same in 
all region of base space, implying that for at least one point they are equal to each other.

\acknowledgments
M. R. was partially supported by Grant No. 2022/45/B/ST2/00013 of the National Science Center, Poland.




\end{document}